\documentclass{article}
\usepackage[utf8]{inputenc}
\usepackage{amsmath, amssymb}
\usepackage{geometry}
\usepackage{graphicx}
\usepackage{float}
\usepackage[colorlinks=true, linkcolor=blue, citecolor=blue, urlcolor=blue]{hyperref}
\title{\textbf{PSDWII: Physical-Structure-Driven Waveform Inversion Imaging}}
\author{CHEN Shengchang\\
        School of Earth Sciences, Zhejiang University, Hangzhou, China\\
        \texttt{chenshengc@zju.edu.cn}}
\date{}

\begin{document}

\maketitle

\begin{abstract}\normalsize
Full-waveform inversion (FWI), stratigraphic physical properties imaging, and stratigraphic structures imaging are the principal techniques in current seismic waveform inversion. These methods are conventionally built upon mathematical optimization theory, which introduces inherent difficulties such as ambiguous physical interpretation of gradient-based updates, aggravated multi-parameter cross-coupling, and prohibitive computational costs. This paper proposes a Physical-Structure-Driven Waveform Inversion Imaging framework (PSDWII), grounded in the recognition that surface-recorded seismic data are directly related to subsurface virtual sources rather than to the model parameters themselves. Based on this recognition, we decompose seismic wave propagation into a three-step physical process: source excitation and incident wave propagation; interaction of the incident wave with heterogeneities, which generates virtual sources with distinct radiation patterns and excites secondary waves; and propagation and reception of the secondary waves. The adjoint of the propagation operator (reverse-time extrapolation) is then used to formulate a linear inversion projection for the virtual sources. By distinguishing different virtual-source expressions and their underlying physical mechanisms, we establish a unified mathematical representation for three inversion tasks: FWI, stratigraphic physical properties imaging, and stratigraphic structures imaging. Deconvolution and linear inversion are respectively adopted to achieve multi-parameter decoupling in FWI and in angle-domain common-image gather inversion for stratigraphic physical properties imaging. The PSDWII framework shifts the conventional ``mathematical optimization driven'' paradigm of waveform inversion to a ``physical-structure-driven'' one. It neither constructs objective functions nor computes their gradients or Hessian inverses. Within this framework, we develop three specific methods: physical-structure-driven FWI (PSDFWI), physical-structure-driven stratigraphic physical properties imaging with angle-domain common-image gather inversion (PSDSI), and physical-structure-driven stratigraphic structures imaging (PSDMig).

\end{abstract}

\section{Introduction}

Current seismic waveform inversion imaging in hydrocarbon exploration and development mainly comprises full-waveform inversion (FWI), stratigraphic physical properties (impedance) imaging, and stratigraphic structures imaging. These techniques have achieved considerable success, yet they also face substantial challenges. The primary issue in current FWI is its immense computational cost, particularly in multi-parameter inversion \cite{tarantola1984, virieux2009, mètivier2024}. This high cost stems from two factors: (1) FWI built upon mathematical optimization does not explicitly account for the physical process of seismic wave propagation; and (2) the physical meaning of the inversion algorithm constructed via local optimization combined with the adjoint-state method remains obscure.

A closer examination reveals that the fundamental difficulty of gradient-based multi-parameter FWI lies in the inherent deficiency of its core operation. The gradient method computes the adjoint wavefield via the adjoint-state method and then cross-correlates (multiplies) it with the incident wavefield and the radiation pattern to obtain the gradient of the objective function. However, according to the physical process of seismic wave propagation, the virtual source is itself the product of the incident wavefield, the radiation pattern, and the model perturbation. The cross-correlation operation in the gradient method superimposes an additional level of mathematical coupling onto the already existing physical coupling inherent in wave propagation. Consequently, the objective function gradient in multi-parameter inversion contains both physical coupling and cross-correlation-induced mathematical coupling. To resolve these couplings, one must resort to the computationally expensive Newton's method, which uses the inverse Hessian for decoupling---essentially a ``pollute-first, remediate-later'' compensatory strategy rather than a solution rooted in physical mechanisms.

Current stratigraphic physical properties imaging and its angle-domain common-image gather inversion are mainly implemented through true-amplitude migration \cite{aki1980} and impedance inversion based on the Zoeppritz equations and their approximations \cite{zhang2013, whitcombe2002, lu2015}. In true-amplitude migration, wavefield extrapolation is based on the wave equation for heterogeneous media, whereas the imaging formula can be regarded as a simplified form of the Zoeppritz equations. The Zoeppritz equations and their approximations describe seismic wave propagation in horizontally layered media and are therefore not suitable for realistic seismic geologic settings. The root causes are twofold: first, appropriate reflection wave equations are lacking for current stratigraphic physical properties imaging and its common-image gather inversion; second, both are constructed upon mathematical optimization methods.

Current stratigraphic structures imaging is built on Claerbout's migration principle, where wavefield extrapolation originates from the wave equation for heterogeneous media, while the imaging formula derives from a simplified version of the Zoeppritz equations for one-dimensional layered media \cite{claerbout1971}. This theoretical gap between the two equations constitutes a deficiency in the theory of stratigraphic structures imaging methods. Again, the root causes are the absence of suitable reflection wave equations for stratigraphic structures imaging and the reliance on mathematical optimization, as in least-squares migration.

To overcome these issues, we must establish a methodological and theoretical framework different from the conventional one. Mathematically, we should move away from general optimization theories (e.g., local optimization with adjoint-state method) and instead tailor methods to the physical structure inherent in the seismic waveform inversion imaging problem. In terms of mathematical physics, we should adopt scattering and reflection wave equations for heterogeneous media that are appropriate for realistic seismic geologic conditions and for waveform inversion imaging, because the general wave equation cannot explicitly describe the propagation of scattered and reflected waves \cite{stolt2012}. Methodologically, we should establish a physical-structure-driven research paradigm based on the specific physical processes of seismic wave (scattering and reflection) propagation, thereby fundamentally avoiding the aforementioned problems.

The FWI, stratigraphic physical properties imaging, and stratigraphic structures imaging presented in this paper constitute a hierarchical system of methodological theories. FWI is a nonlinear inversion imaging method that fully utilizes seismic waveform information. Its ultimate goal is to provide high-resolution, high-fidelity inversion results of the subsurface medium; its minimum goal is to provide a low-resolution inversion result with accurate kinematic characteristics (a smooth subsurface model) for subsequent stratigraphic physical properties imaging and stratigraphic structures imaging. If FWI achieves its ultimate goal, the latter two become essentially unnecessary. If FWI can only provide a kinematically accurate low-resolution result, then stratigraphic physical properties imaging and stratigraphic structures imaging are required to obtain information on subsurface property variations and structural information. Both are linear inversion imaging of primary reflections, conducted under the condition of a smooth subsurface model with accurate kinematic characteristics. Stratigraphic physical properties imaging consists of two components: fast imaging of property variations and fine imaging of lithologic parameters through angle-domain common-image gather waveform inversion. Stratigraphic structures imaging achieves structural images by imaging the local reflectivity of the boundaries of subsurface heterogeneities (including strata).

The main contributions of this paper are:

\begin{enumerate}
\item Proposing a physical-structure-driven seismic waveform inversion imaging framework (PSDWII): treating the seismic wave propagation process as a system with a specific physical structure, namely, system input (source excitation and incident wave propagation) $\rightarrow$ system response (interaction of incident wave with heterogeneities generating virtual sources with radiation patterns) $\rightarrow$ system output (virtual sources exciting secondary waves that propagate to receivers and are recorded). Through reverse-time (adjoint) propagation of the system output (observed wavefield), an approximate inversion of the system response (virtual source) is obtained. Combined with the mechanism and theoretical expression of the virtual source and the system input (incident wave), the PSDWII framework is constructed.

\item Interpreting the reverse-time extrapolated wavefield (adjoint wavefield) of the observed data as the approximate inversion of the subsurface virtual source, thereby endowing the adjoint wavefield obtained from the adjoint-state method with a clear physical meaning.

\item Proposing the physical-structure-driven FWI method (PSDFWI): starting from the perturbational form of the wave equation---the scattering wave equation---and utilizing the approximate inversion result of the scattering virtual source obtained by reverse-time propagation of the wavefield residual, combined with the mechanism and theoretical expression of the scattering virtual source and the incident wavefield, we construct the physical-structure-driven FWI method. PSDFWI is a stepwise inversion method with clear physical concepts, computational simplicity, and high efficiency, particularly suitable for multi-parameter FWI.

\item Proposing the physical-structure-driven stratigraphic physical properties imaging and angle-domain common-image gather waveform linear inversion method (PSDSI): starting from the wave equation, we construct the primary reflection wave equation for body-reflection virtual sources with stratigraphic physical property parameter relative perturbations as variables. Using the approximate inversion result of the body-reflection virtual source obtained by reverse-time propagation of the recorded primary reflection wavefield, combined with the mechanism and theoretical expression of the primary body-reflection virtual source and the incident wavefield, we establish a fully wave-equation-based stratigraphic physical properties imaging method and a common-image gather inversion method independent of the Zoeppritz equations and their approximations.

\item Proposing the physical-structure-driven stratigraphic structures imaging (migration) method (PSDMig): based on the body-reflection virtual source, we define the directional derivative of the relative perturbation of stratum model parameters along the incident wave propagation direction as the local reflectivity of the stratigraphic boundary. We construct the surface-reflection virtual source and primary reflection wave equation with the local reflectivity of the stratigraphic boundary as the variable. Using the approximate inversion result of the surface-reflection virtual source obtained by reverse-time propagation of the recorded primary reflection wavefield, combined with the mechanism and theoretical expression of the surface-reflection virtual source and the incident wavefield, we establish a fully wave-equation-based stratigraphic structures imaging method (PSDMig)---a migration method different from the concept of Claerbout's migration imaging.

\item Realizing the transformation of the seismic waveform inversion imaging research paradigm from ``mathematical optimization driven'' to ``physical-structure-driven.'' This transforms seismic waveform inversion imaging research from relying on general mathematical optimization methods---with poor physical interpretability, high computational complexity, and low computational efficiency---to tailoring physical-structure-driven methods according to the physical structure of the problem---with strong physical interpretability and computational conciseness and efficiency.
\end{enumerate}

The core steps of the three methods proposed in this paper have been validated in our previous numerical experiments \cite{chen2016a, chen2016b, liu2020, chen2022a, chen2022b}. This paper focuses on elaborating their theoretical foundations and unified framework.

\section{Physical Process and Physical Structure of Seismic Wave Propagation}

For the common surface-source and surface-receiver acquisition system in seismic exploration, the seismic wave propagation process is as follows: the surface source excites incident waves, which propagate and encounter heterogeneities, generating seismic virtual sources with radiation patterns. These virtual sources excite secondary waves (scattered and reflected waves), which propagate to the surface and are recorded by receivers. The physical process of seismic wave propagation can be summarized as ``two propagations'' + ``one virtual source'', i.e., incident wavefield propagation and secondary wavefield propagation + seismic virtual sources with radiation patterns. This physical process can also be viewed as a physical system process: system input (source excitation + incident wave propagation) $\rightarrow$ system response (interaction of incident wave with heterogeneities generating virtual sources + virtual sources exciting secondary waves) $\rightarrow$ system output (secondary wave propagation + receiver recording).

The physical structure of seismic wave propagation: two propagations + one virtual source, i.e., the physical structure of system input (incident wavefield) $\rightarrow$ system response (subsurface virtual source) $\rightarrow$ system output (observed wavefield).

For the general wave equation,
\begin{equation}
\mathrm{L}(\mathbf{m})\mathbf{u} = \mathbf{s}. \tag{1}
\label{eq:general_wave}
\end{equation}
where \(\mathrm{L}\) denotes the wave operator; \(\mathbf{m}\) denotes the subsurface medium model parameters; \(\mathbf{u}\) denotes the seismic wavefield; and \(\mathbf{s}\) denotes the source function. Given a background model \(\mathbf{m}_0\), the corresponding background wavefield \(\mathbf{u}_0\) satisfies
\begin{equation}
\mathrm{L}(\mathbf{m}_0)\mathbf{u}_0 = \mathbf{s}. \tag{2}
\label{eq:background_wave}
\end{equation}
Perturbing \(\mathbf{m}\) and \(\mathbf{u}\) as \(\mathbf{m} = \mathbf{m}_0 + \delta\mathbf{m}\) and \(\mathbf{u} = \mathbf{u}_0 + \delta\mathbf{u}\), where \(\delta\mathbf{m}\) is the model perturbation and \(\delta\mathbf{u}\) is the perturbed wavefield. Substituting the perturbed model and wavefield into Eq.~\eqref{eq:general_wave} and combining with Eq.~\eqref{eq:background_wave}, the perturbed wavefield equation is obtained:
\begin{equation}
\mathrm{L}(\mathbf{m}_0)\delta\mathbf{u} = \mathbf{V}_{\mathrm{s}}. \tag{3}
\label{eq:perturbed_wave}
\end{equation}
where \(\mathbf{V}_{\mathrm{s}}\) is the virtual source generating the perturbed (secondary) wavefield, given by
\begin{equation}
\mathbf{V}_{\mathrm{s}} = -\left.\frac{\partial \mathrm{L}(\mathbf{m})}{\partial \mathbf{m}}\right|_{\mathbf{m} = \mathbf{m}_0} \delta\mathbf{m}(\mathbf{u}_0 + \delta\mathbf{u}). \tag{4}
\label{eq:virtual_source_full}
\end{equation}
where \(\frac{\partial \mathrm{L}(\mathbf{m})}{\partial \mathbf{m}}\) is the derivative of the wave operator with respect to the model parameters (abbreviated as the wave operator derivative), which primarily determines the radiation pattern of the virtual source. \(\mathbf{V}_{\mathrm{s}}\) is nonlinearly related to the model perturbation \(\delta\mathbf{m}\), and is therefore called the nonlinear virtual source (full-wave virtual source). If the perturbed wavefield is much smaller than the background wavefield, i.e., \(\delta\mathbf{u} \ll \mathbf{u}_0\), then \(\mathbf{V}_{\mathrm{s}}\) reduces to the primary wave virtual source \(\mathbf{V}_{\mathrm{s}}^{\mathrm{p}}\), which is linearly related to \(\delta\mathbf{m}\):
\begin{equation}
\mathbf{V}_{\mathrm{s}}^{\mathrm{p}} = -\left.\frac{\partial \mathrm{L}(\mathbf{m})}{\partial \mathbf{m}}\right|_{\mathbf{m} = \mathbf{m}_0} \delta\mathbf{m}\mathbf{u}_0. \tag{5}
\label{eq:virtual_source_primary}
\end{equation}

Using the Green's function (also called the wavefield propagation operator) of the wave equation Eq.~\eqref{eq:background_wave}, the background wavefield (also called the incident wavefield) in Eq.~\eqref{eq:background_wave} and the perturbed wavefield in Eq.~\eqref{eq:perturbed_wave} can be respectively expressed as:
\begin{equation}
\mathbf{u}_0 = \mathrm{L}^{-1}(\mathbf{m}_0)\mathbf{s} = \mathbf{G}_{\mathrm{i}}(\mathbf{m}_0)\mathbf{s}. \tag{6}
\label{eq:incident_wave}
\end{equation}
\begin{equation}
\delta\mathbf{u} = \mathrm{L}^{-1}(\mathbf{m}_0)\mathbf{V}_{\mathrm{s}} = \mathbf{G}_{\mathrm{s}}(\mathbf{m}_0)\mathbf{V}_{\mathrm{s}}. \tag{7}
\label{eq:secondary_wave}
\end{equation}
where \(\mathbf{G}_{\mathrm{i}}(\mathbf{m}_0)\) and \(\mathbf{G}_{\mathrm{s}}(\mathbf{m}_0)\) denote the incident wave propagation operator and secondary wave propagation operator, respectively. The corresponding primary secondary wave equation and primary secondary wavefield are:
\begin{equation}
\mathrm{L}(\mathbf{m}_0)\delta\mathbf{u}^{\mathrm{p}} = \mathbf{V}_{\mathrm{s}}^{\mathrm{p}}. \tag{8}
\label{eq:primary_secondary_wave_eq}
\end{equation}
\begin{equation}
\delta\mathbf{u}^{\mathrm{p}} = \mathrm{L}^{-1}(\mathbf{m}_0)\mathbf{V}_{\mathrm{s}}^{\mathrm{p}} = \mathbf{G}_{\mathrm{s}}(\mathbf{m}_0)\mathbf{V}_{\mathrm{s}}^{\mathrm{p}}. \tag{9}
\label{eq:primary_secondary_wave}
\end{equation}

From the above, Eqs.~\eqref{eq:incident_wave}, \eqref{eq:secondary_wave}, and \eqref{eq:primary_secondary_wave} represent the incident wave propagation and secondary wave propagation in the physical structure of seismic wave propagation, while Eqs.~\eqref{eq:virtual_source_full} and \eqref{eq:virtual_source_primary} represent the nonlinear virtual source and primary (linear) virtual source generated by the incident wavefield acting on the model perturbation \(\delta\mathbf{m}\).

The type of secondary virtual source generated by the incident wavefield acting on heterogeneities depends on the relationship between the heterogeneity scale \(a\) and the dominant wavelength \(\lambda\) of the seismic wave \cite{robein2010}. If the scale \(a\) of the heterogeneity  \(\delta\mathbf{m}\) is less than or equal to the dominant wavelength \(\lambda\), i.e., \(a \le \lambda\), then the virtual source \(\mathbf{V}_{\mathrm{s}}\) (\(\mathbf{V}_{\mathrm{s}}^{\mathrm{p}}\)) is a scattering virtual source generating scattered waves (also called a volume scattering virtual source), denoted as \(\mathbf{V}_{\mathrm{ss}}(\mathbf{m}_0,\delta\mathbf{m})\) (\(\mathbf{V}_{\mathrm{ss}}^{\mathrm{p}}(\mathbf{m}_0,\delta\mathbf{m})\)). If the heterogeneity scale \(a\) is greater than the dominant wavelength \(\lambda\), i.e., \(a > \lambda\), and considering the interference effect of seismic waves, then the virtual source \(\mathbf{V}_{\mathrm{s}}\) (\(\mathbf{V}_{\mathrm{s}}^{\mathrm{p}}\)) is a reflection virtual source generating reflected waves (also called a volume reflection virtual source), denoted as \(\mathbf{V}_{\mathrm{sr}}(\mathbf{m}_0,\delta\mathbf{m},\theta)\) (\(\mathbf{V}_{\mathrm{sr}}^{\mathrm{p}}(\mathbf{m}_0,\delta\mathbf{m},\theta)\)), where \(\theta\) is the reflection opening angle between the reflected and incident waves. Let \(\alpha\) be the relative perturbation between the perturbed and background models, i.e., \(\alpha = \frac{\delta\mathbf{m}}{\mathbf{m}_0}\). Then the volume reflection virtual source with \(\alpha\) as the model parameter is denoted as \(\mathbf{V}_{\mathrm{sr}}(\mathbf{m}_0,\alpha,\theta)\) (\(\mathbf{V}_{\mathrm{sr}}^{\mathrm{p}}(\mathbf{m}_0,\alpha,\theta)\)). To consider the reflection virtual source with the local reflectivity \(r\) of the tangent plane of the reflector boundary as the model parameter (also called the surface reflection virtual source), we define the local reflectivity \(r\) of the tangent plane of the reflector boundary as the directional derivative of the model parameter perturbation \(\alpha\) along the incident wave propagation direction \cite{stolt2012, chen2016a, bleistein2001}:
\begin{equation}
r = \frac{\partial\alpha}{\partial \mathbf{J}}. \tag{10}
\label{eq:local_reflectivity}
\end{equation}
The wavenumber-domain form of Eq.~\eqref{eq:local_reflectivity} is:
\begin{equation}
r = i\mathbf{k}_{\mathrm{j}}(\alpha). \tag{11}
\label{eq:local_reflectivity_wavenumber}
\end{equation}
where \(\mathbf{J}\) is the incident wave propagation direction; \(i\) is the imaginary unit; \(\mathbf{k}_{\mathrm{j}}\) is the local wavenumber vector in the incident wave propagation direction, with \(\mathbf{k}_{\mathrm{j}} = \frac{\omega}{\mathbf{V}_0}\vec{\mathbf{e}}_{\mathrm{j}}\); \(\omega\) is the angular frequency; \(\mathbf{V}_0\) is the velocity in the background model; \(\vec{\mathbf{e}}_{\mathrm{j}}\) is the unit vector in the incident wave propagation direction; \(r\) is also called the local reflectivity of the reflector boundary in the incident wave propagation direction. The surface reflection virtual source with \(r\) as the model parameter is denoted as \(\mathbf{V}_{\mathrm{sr}}(\mathbf{m}_0,r)\) (\(\mathbf{V}_{\mathrm{sr}}^{\mathrm{p}}(\mathbf{m}_0,r)\)).

For the acoustic wave equation with density \(\rho(\mathbf{x})\) and velocity \(\mathbf{v}(\mathbf{x})\) as model parameters:
\begin{equation}
\left\{\frac{1}{\rho(\mathbf{x})\mathbf{v}^2(\mathbf{x})}\frac{\partial^2}{\partial t^2} - \nabla \cdot \left[\frac{1}{\rho(\mathbf{x})}\nabla\right]\right\} \mathbf{u}(\mathbf{x},t;\mathbf{x}_s) = \mathbf{s}(\mathbf{x}_s,t). \tag{12}
\label{eq:acoustic_wave}
\end{equation}
where \(\mathbf{x}\) is the spatial coordinate; \(t\) is time; and \(\mathbf{x}_s\) is the source point spatial coordinate. Denote \(\rho(\mathbf{x})\), \(\mathbf{v}(\mathbf{x})\), \(\mathbf{u}(\mathbf{x},t;\mathbf{x}_s)\), and \(\mathbf{s}(\mathbf{x}_s,t)\) as \(\rho\), \(\mathbf{v}\), \(\mathbf{u}\), and \(\mathbf{s}\) for brevity. Given the initial models \(\rho_0\) and \(\mathbf{v}_0\), with \(\rho = \rho_0 + \delta\rho\) and \(\mathbf{v} = \mathbf{v}_0 + \delta\mathbf{v}\), the corresponding specific expressions for the scattering virtual source, volume reflection virtual source, and surface reflection virtual source are:
\begin{equation}
\mathbf{V}_{\mathrm{ss}}(\mathbf{v}_0,\rho_0,\delta\mathbf{v},\delta\rho) = \frac{1}{\rho_0\mathbf{v}_0^2} \left(\frac{\delta\rho}{\rho_0} + \frac{2\delta\mathbf{v}}{\mathbf{v}_0}\right)\frac{\partial^2\mathbf{u}}{\partial t^2} - \nabla \cdot \left(\frac{\delta\rho}{\rho_0^2}\nabla\mathbf{u}\right). \tag{13}
\label{eq:scattering_virtual_source}
\end{equation}
\begin{equation}
\mathbf{V}_{\mathrm{ss}}^{\mathrm{p}}(\mathbf{v}_0,\rho_0,\delta\mathbf{v},\delta\rho) = \frac{1}{\rho_0\mathbf{v}_0^2} \left(\frac{\delta\rho}{\rho_0} + \frac{2\delta\mathbf{v}}{\mathbf{v}_0}\right)\frac{\partial^2\mathbf{u}_0}{\partial t^2} - \nabla \cdot \left(\frac{\delta\rho}{\rho_0^2}\nabla\mathbf{u}_0\right). \tag{14}
\label{eq:scattering_virtual_source_primary}
\end{equation}
\begin{equation}
\mathbf{V}_{\mathrm{sr}}(\mathbf{v}_0,\rho_0,\alpha_{\mathrm{v}},\alpha_{\rho},\theta) = \frac{(1 + \cos\theta)\alpha_{\rho} + \alpha_{\mathrm{v}}}{\rho_0\mathbf{v}_0^2}\frac{\partial^2\mathbf{u}}{\partial t^2} = \frac{\mathbf{I}_{\mathrm{r}}(\alpha_{\rho},\alpha_{\mathrm{v}},\theta)}{\rho_0\mathbf{v}_0^2}\frac{\partial^2\mathbf{u}}{\partial t^2}. \tag{15}
\label{eq:volume_reflection_virtual_source}
\end{equation}
\begin{equation}
\mathbf{V}_{\mathrm{sr}}^{\mathrm{p}}(\mathbf{v}_0,\rho_0,\alpha_{\mathrm{v}},\alpha_{\rho},\theta) = \frac{\mathbf{I}_{\mathrm{r}}(\alpha_{\rho},\alpha_{\mathrm{v}},\theta)}{\rho_0\mathbf{v}_0^2}\frac{\partial^2\mathbf{u}_0}{\partial t^2}. \tag{16}
\label{eq:volume_reflection_virtual_source_primary}
\end{equation}
\begin{equation}
\mathbf{V}_{\mathrm{sr}}(\mathbf{v}_0,\rho_0,r) = -\frac{r}{\rho_0\mathbf{v}_0}\frac{\partial\mathbf{u}}{\partial t}. \tag{17}
\label{eq:surface_reflection_virtual_source}
\end{equation}
\begin{equation}
\mathbf{V}_{\mathrm{sr}}^{\mathrm{p}}(\mathbf{v}_0,\rho_0,r) = -\frac{r}{\rho_0\mathbf{v}_0}\frac{\partial\mathbf{u}_0}{\partial t}. \tag{18}
\label{eq:surface_reflection_virtual_source_primary}
\end{equation}
In the above equations, \(\alpha_{\rho} = \frac{\delta\rho}{\rho_0}\); \(\alpha_{\mathrm{v}} = \frac{2\delta\mathbf{v}}{\mathbf{v}_0}\); \(\mathbf{I}_{\mathrm{r}}(\alpha_{\rho},\alpha_{\mathrm{v}},\theta) = (1 + \cos\theta)\alpha_{\rho} + \alpha_{\mathrm{v}}\) is called the relative perturbation of the reflector's impedance; \(r = \frac{i\omega}{\mathbf{v}_0}(\mathbf{I}_{\mathrm{r}}(\alpha_{\rho},\alpha_{\mathrm{v}},\theta))\bar{\mathbf{e}}_{\mathrm{j}}\).

From the above physical process and physical structure of seismic wave propagation, it can be seen that both scattered and reflected waves share the same incident wave propagation and secondary wave (scattered and reflected wave) propagation, differing only in the virtual source.

\section{Physical-Structure-Driven Seismic Waveform Inversion Imaging Framework (PSDWII)}

Based on the physical process and physical structure of seismic wave propagation, the following three observations can be made: (1) the observed wavefield is primarily related to the virtual source generating the secondary waves and the propagation of the secondary waves; (2) the virtual source with radiation patterns is the result of the joint action of the incident wavefield, the derivative of the wave operator with respect to the model parameters, and the heterogeneity---this action is a time-consistent local action, and the radiation pattern of the virtual source is primarily determined by the wave operator derivative; (3) the source and the incident wave propagation from the source to the heterogeneity are both contained in the incident wavefield of the virtual source. These three observations are the starting points for the construction of the PSDWII framework.

Based on the above observations, we consider that the inversion imaging of seismic waveforms should first be the inversion imaging of subsurface virtual sources, i.e., inverting the subsurface virtual source from the observed wavefield; then, based on the theoretical expression of the subsurface virtual source and the time-consistent local action in the virtual source, the incident wavefield and the radiation pattern of the virtual source are removed from the inverted virtual source, achieving the inversion imaging of the target. Different virtual sources generate different secondary wavefields, leading to different inversion targets.

The physical-structure-driven seismic waveform inversion imaging framework:

\textbf{Step 1: Using the adjoint operator \(\mathbf{G}_{\mathrm{s}}^{*}(\mathbf{m}_0)\) of the secondary wave propagation operator to obtain the approximate inversion result of the subsurface virtual source from the observed secondary wavefield.} From Eqs.~\eqref{eq:secondary_wave} and \eqref{eq:primary_secondary_wave}, the approximate inversion results \(\tilde{\mathbf{V}}_{\mathrm{s}}\) and \(\tilde{\mathbf{V}}_{\mathrm{s}}^{\mathrm{p}}\) of the full-wave virtual source and primary virtual source can be obtained:
\begin{equation}
\tilde{\mathbf{V}}_{\mathrm{s}} = \mathbf{G}_{\mathrm{s}}^{*}(\mathbf{m}_0)\delta\mathbf{u}. \tag{19}
\label{eq:approx_virtual_source_full}
\end{equation}
\begin{equation}
\tilde{\mathbf{V}}_{\mathrm{s}}^{\mathrm{p}} = \mathbf{G}_{\mathrm{s}}^{*}(\mathbf{m}_0)\delta\mathbf{u}^{\mathrm{p}}. \tag{20}
\label{eq:approx_virtual_source_primary}
\end{equation}
The specific computation of Eqs.~\eqref{eq:approx_virtual_source_full} and \eqref{eq:approx_virtual_source_primary} is implemented through reverse-time (adjoint) propagation of the wavefield. The obtained approximate inversion result of the virtual source is exactly the so-called adjoint wavefield. Therefore, the adjoint wavefield can be interpreted as the approximate inversion of the subsurface virtual source---this is the physical meaning of the adjoint wavefield.

\textbf{Step 2: Based on the theoretical expression of the subsurface virtual source and the time-consistent local action in the virtual source, remove the incident wavefield and the radiation pattern of the virtual source from the inversion imaging result of the subsurface virtual source, achieving the inversion imaging of the target.} The incident wavefield is obtained by forward modeling of the wavefield, and the radiation pattern of the virtual source is primarily determined by the derivative of the wave operator with respect to the inversion target (model parameters). Based on the theoretical expression of the full-wave virtual source Eq.~\eqref{eq:virtual_source_full} and the approximate inversion result \(\tilde{\mathbf{V}}_{\mathrm{s}}\), the approximate solution for the model parameter perturbation \(\delta\mathbf{m}\) can be obtained:
\begin{equation}
\delta\mathbf{m} = -\tilde{\mathbf{V}}_{\mathrm{s}} \bigg/ \frac{\partial \mathrm{L}(\mathbf{m}_0)}{\partial \mathbf{m}}(\mathbf{u}_0 + \delta\mathbf{u}). \tag{21}
\label{eq:nonlinear_inversion}
\end{equation}
Since the perturbed wavefield \(\delta\mathbf{u}\) on the right-hand side of Eq.~\eqref{eq:nonlinear_inversion} is related to the model perturbation \(\delta\mathbf{m}\) to be determined, the solution of Eq.~\eqref{eq:nonlinear_inversion} is nonlinear and requires linearization:
\begin{equation}
\delta\mathbf{m} = -\tilde{\mathbf{V}}_{\mathrm{s}} \bigg/ \frac{\partial \mathrm{L}(\mathbf{m}_0)}{\partial \mathbf{m}}\mathbf{u}_0. \tag{22}
\label{eq:linearized_inversion}
\end{equation}
Based on the theoretical expression of the primary virtual source Eq.~\eqref{eq:virtual_source_primary} and the approximate inversion result \(\tilde{\mathbf{V}}_{\mathrm{s}}^{\mathrm{p}}\), the approximate solution for the model parameter perturbation \(\delta\mathbf{m}\) can be obtained:
\begin{equation}
\delta\mathbf{m} = -\tilde{\mathbf{V}}_{\mathrm{s}}^{\mathrm{p}} \bigg/ \frac{\partial \mathrm{L}(\mathbf{m}_0)}{\partial \mathbf{m}}\mathbf{u}_0. \tag{23}
\label{eq:primary_linear_inversion}
\end{equation}
Equation~\eqref{eq:primary_linear_inversion} represents a linear solution based on a linear expression, because \(\tilde{\mathbf{V}}_{\mathrm{s}}^{\mathrm{p}}\) is linearly related to \(\delta\mathbf{m}\). Equation~\eqref{eq:linearized_inversion} represents a linear solution under linearization of the nonlinear problem. The operations in Eqs.~\eqref{eq:nonlinear_inversion}, \eqref{eq:linearized_inversion}, and \eqref{eq:primary_linear_inversion} are deconvolution operations, which require that the virtual source obtained by reverse-time extrapolation is time-consistent with the incident wavefield. This time consistency is also a necessary condition for determining whether the ``virtual source'' obtained by reverse-time extrapolation is the true virtual source generating the secondary wavefield.

If the model parameter \(\mathbf{m}\) contains multiple parameters, i.e., \(\mathbf{m} = (m_1, m_2, \ldots, m_N)^{\mathrm{T}}\), then the full-wave virtual source in Eq.~\eqref{eq:virtual_source_full} and the primary virtual source in Eq.~\eqref{eq:virtual_source_primary} can be respectively written as:
\begin{equation}
\mathbf{V}_{\mathrm{s}} = -\sum_{j=1}^{N}\frac{\partial \mathrm{L}(\mathbf{m}_0)}{\partial m_j}\delta m_j(\mathbf{u}_0 + \delta\mathbf{u}) \approx -\sum_{j=1}^{N}\frac{\partial \mathrm{L}(\mathbf{m}_0)}{\partial m_j}\delta m_j\mathbf{u}_0. \tag{24}
\label{eq:multi_virtual_source_full}
\end{equation}
\begin{equation}
\mathbf{V}_{\mathrm{s}}^{\mathrm{p}} = -\sum_{j=1}^{N}\frac{\partial \mathrm{L}(\mathbf{m}_0)}{\partial m_j}\delta m_j\mathbf{u}_0. \tag{25}
\label{eq:multi_virtual_source_primary}
\end{equation}
From Eqs.~\eqref{eq:linearized_inversion} and \eqref{eq:primary_linear_inversion}, the update \(\delta m_i\) for a single model parameter \(m_i\) can be obtained:
\begin{equation}
\delta m_i = -\tilde{\mathbf{V}}_{\mathrm{s}} \bigg/ \frac{\partial \mathrm{L}(\mathbf{m}_0)}{\partial m_i}\mathbf{u}_0. \tag{26}
\label{eq:single_correction_nonlinear}
\end{equation}
\begin{equation}
\delta m_i = -\tilde{\mathbf{V}}_{\mathrm{s}}^{\mathrm{p}} \bigg/ \frac{\partial \mathrm{L}(\mathbf{m}_0)}{\partial m_i}\mathbf{u}_0. \tag{27}
\label{eq:single_correction_primary}
\end{equation}

To understand the influence of cross-talk caused by multi-parameter coupling on the inversion, substituting the approximations of Eqs.~\eqref{eq:secondary_wave}, \eqref{eq:approx_virtual_source_full}, and \eqref{eq:multi_virtual_source_full} into Eq.~\eqref{eq:single_correction_nonlinear} yields:
\begin{equation}
\delta m_i = \frac{\mathbf{G}_{\mathrm{s}}^{*}(\mathbf{m}_0)\mathbf{G}_{\mathrm{s}}(\mathbf{m}_0)\sum_{j=1}^{N}\frac{\partial \mathrm{L}(\mathbf{m}_0)}{\partial m_j}\delta m_j\mathbf{u}_0}{\frac{\partial \mathrm{L}(\mathbf{m}_0)}{\partial m_i}\mathbf{u}_0}
= \frac{\mathbf{G}_{\mathrm{s}}^{*}(\mathbf{m}_0)\mathbf{G}_{\mathrm{s}}(\mathbf{m}_0)\mathbf{u}_0}{\mathbf{u}_0}(\delta m_i + \mathbf{ct}). \tag{28}
\label{eq:cross_talk}
\end{equation}
where \(\mathbf{ct}\) represents the cross-talk caused by multi-parameter coupling:
\begin{equation}
\mathbf{ct} = \sum_{j=1, j\neq i}^{N} \left( \frac{\partial \mathrm{L}(\mathbf{m}_0)}{\partial m_j}\delta m_j \bigg/ \frac{\partial \mathrm{L}(\mathbf{m}_0)}{\partial m_i} \right). \tag{29}
\label{eq:ct_definition}
\end{equation}

\textbf{Remark on the incident wavefield in Eq.~\eqref{eq:cross_talk}:} The incident wavefield \(\mathbf{u}_0\) appears in both the numerator and denominator on the right-hand side of Eq.~\eqref{eq:cross_talk}. In the numerator, \(\mathbf{u}_0\) propagates from the model perturbation location to the receivers via \(\mathbf{G}_{\mathrm{s}}\), and then back to the model perturbation location via \(\mathbf{G}_{\mathrm{s}}^{*}\). In the denominator, \(\mathbf{u}_0\) is the in-situ incident wavefield at the model perturbation location.

In actual computation, the \(\mathbf{u}_0\) in the denominator is usually not the in-situ incident wavefield, but the \textbf{reconstructed incident wavefield} obtained by propagating from the model perturbation location to the computational boundary via \(\mathbf{G}_{\mathrm{i}}\), and then back to the model perturbation location via \(\mathbf{G}_{\mathrm{i}}^{*}\), i.e., \(\tilde{u}_0 = \mathbf{G}_{\mathrm{i}}^{*}\mathbf{G}_{\mathrm{i}}\mathbf{u}_0\). This is a common incident wavefield reconstruction method used in seismic waveform inversion imaging.

When the reconstructed incident wavefield is used in the denominator, although the \(\mathbf{G}_{\mathrm{s}}^{*}\mathbf{G}_{\mathrm{s}}\mathbf{u}_0\) in the numerator and the \(\mathbf{G}_{\mathrm{i}}^{*}\mathbf{G}_{\mathrm{i}}\mathbf{u}_0\) in the denominator have different propagation paths, their propagation times are consistent. Under the condition of time consistency, the propagation operators and adjoint propagation operators that appear in both the numerator and denominator can be approximately cancelled, i.e.,
\begin{equation}
\frac{\mathbf{G}_{\mathrm{s}}^{*}(\mathbf{m}_0)\mathbf{G}_{\mathrm{s}}(\mathbf{m}_0)\mathbf{u}_0}{{u}_0} \approx\frac{\mathbf{G}_{\mathrm{s}}^{*}(\mathbf{m}_0)\mathbf{G}_{\mathrm{s}}(\mathbf{m}_0)\mathbf{u}_0}{\tilde{u}_0} =\frac{\mathbf{G}_{\mathrm{s}}^{*}(\mathbf{m}_0)\mathbf{G}_{\mathrm{s}}(\mathbf{m}_0)\mathbf{u}_0}{\mathbf{G}_{\mathrm{i}}^{*}(\mathbf{m}_0)\mathbf{G}_{\mathrm{i}}(\mathbf{m}_0)\mathbf{u}_0} \approx 1. \tag{30}
\label{eq:approx_cancellation}
\end{equation}
The approximation Eq.~\eqref{eq:approx_cancellation} above indicates that if the incident wavefield in the denominator of Eq.~\eqref{eq:cross_talk} is the reconstructed incident wavefield, then Eq.~\eqref{eq:cross_talk} has good fidelity, implying that the deconvolution in the PSDWII framework has good amplitude preservation. This recognition has not been fully appreciated in current seismic waveform inversion imaging research.

For multi-parameter inversion, the inversion in Eq.~\eqref{eq:single_correction_nonlinear} is mainly affected by four factors: (1) the linearization approximation in Eq.~\eqref{eq:multi_virtual_source_full}; (2) the fact that \(\tilde{\mathbf{V}}_{\mathrm{s}}\) is obtained using the adjoint operator rather than the inverse operator; (3) the cross-talk \(\mathbf{ct}\) between multiple parameters; and (4) the stability of the deconvolution in Eq.~\eqref{eq:single_correction_nonlinear}. The inversion in Eq.~\eqref{eq:single_correction_primary} is also affected by the last three factors. If the model parameterization is chosen appropriately so that the radiation patterns of different model parameters \(m_i\) and \(m_j\) have poor coherence, then \(\mathbf{ct} \to 0\), i.e., the multi-parameter cross-talk can be eliminated.

PSDWII is a stepwise inversion framework that does not involve objective functions, nor the computation of objective function gradients or Hessian matrices and their inverses. Each step has clear physical meaning and is computationally efficient with low complexity. It is also the unified framework for the full-waveform inversion, stratigraphic physical properties imaging and its angle-domain common-image gather linear inversion, and stratigraphic structures imaging methods developed in this paper.

\section{Physical-Structure-Driven Full-Waveform Inversion Method (PSDFWI)}

According to the PSDWII above, given the initial model \(\mathbf{m}_0\), the computational steps of PSDFWI are:

\begin{enumerate}
\item Compute the theoretical wavefield corresponding to the initial model, i.e., Eq.~\eqref{eq:incident_wave}, which can also be regarded as the incident wavefield;
\item Compute the wavefield residual at the receivers: observed wavefield minus theoretical wavefield;
\item Use the wavefield residual to invert the scattering virtual source, i.e., Eq.~\eqref{eq:approx_virtual_source_full};
\item Remove the incident wavefield and radiation pattern from the inverted virtual source to obtain the correction to the initial model, i.e., Eq.~\eqref{eq:linearized_inversion};
\item Update the initial model using the obtained model correction: \(\mathbf{m}_0 = \mathbf{m}_0 + \delta\mathbf{m}\);
\item Replace the original initial model with the updated one and return to step 1.
\end{enumerate}

For the acoustic wave equation Eq.~\eqref{eq:acoustic_wave} in PSDFWI, given the initial models \(\rho_0\) and \(\mathbf{v}_0\), through steps 1 to 3 above, the inversion result \(\tilde{\mathbf{V}}_{\mathrm{ss}}\) of the subsurface scattering virtual source can be obtained. Then, according to the theoretical expression of the scattering virtual source Eq.~\eqref{eq:scattering_virtual_source} and the linearized inversion formula Eq.~\eqref{eq:single_correction_nonlinear}, the correction formulas for velocity \(\mathbf{v}\) and density \(\rho\) can be obtained:
\begin{equation}
\delta\mathbf{v} = \tilde{\mathbf{V}}_{\mathrm{ss}} \bigg/ \frac{2}{\rho_0\mathbf{v}_0^3}\frac{\partial^2\mathbf{u}_0}{\partial t^2}
= \frac{\rho_0\mathbf{v}_0^3}{2}\tilde{\mathbf{V}}_{\mathrm{ss}} \bigg/ \frac{\partial^2\mathbf{u}_0}{\partial t^2}. \tag{31}
\label{eq:velocity_correction}
\end{equation}
\begin{equation}
\delta\rho = \tilde{\mathbf{V}}_{\mathrm{ss}} \bigg/ \left( \frac{1}{\rho_0^2\mathbf{v}_0^2}\frac{\partial^2\mathbf{u}_0}{\partial t^2} - \nabla \cdot \left(\frac{1}{\rho_0^2}\nabla\mathbf{u}_0\right) \right). \tag{32}
\label{eq:density_correction}
\end{equation}
Using the obtained model parameter corrections, update the initial models: \(\mathbf{v}_0 = \mathbf{v}_0 + \delta\mathbf{v}\), \(\rho_0 = \rho_0 + \delta\rho\), and then iterate.

If the acoustic wave equation Eq.~\eqref{eq:acoustic_wave} is parameterized in terms of bulk modulus \(\kappa(\mathbf{x})\) and density \(\rho(\mathbf{x})\):
\begin{equation}
\left\{\frac{1}{\kappa(\mathbf{x})}\frac{\partial^2}{\partial t^2} - \nabla \cdot \left[\frac{1}{\rho(\mathbf{x})}\nabla\right]\right\} \mathbf{u}(\mathbf{x},\mathbf{x}_s,t) = \mathbf{s}(\mathbf{x}_s,t). \tag{33}
\label{eq:bulk_modulus_wave}
\end{equation}
Given the initial models \(\kappa_0\) and \(\rho_0\), the scattering virtual source for this acoustic wave equation has the following theoretical expression:
\begin{equation}
\mathbf{V}_{\mathrm{ss}} = \frac{\delta\kappa}{\kappa_0^2}\frac{\partial^2\mathbf{u}}{\partial t^2} - \nabla \cdot \left[\frac{\delta\rho}{\rho_0^2}\nabla\mathbf{u}\right]
\approx \frac{\delta\kappa}{\kappa_0^2}\frac{\partial^2\mathbf{u}_0}{\partial t^2} - \nabla \cdot \left[\frac{\delta\rho}{\rho_0^2}\nabla\mathbf{u}_0\right]. \tag{34}
\label{eq:scattering_virtual_source_bulk}
\end{equation}
The correction formulas for \(\kappa\) and \(\rho\) can be obtained:
\begin{equation}
\delta\kappa = \kappa_0^2 \tilde{\mathbf{V}}_{\mathrm{ss}} \bigg/ \frac{\partial^2\mathbf{u}_0}{\partial t^2}. \tag{35}
\label{eq:bulk_correction}
\end{equation}
\begin{equation}
\delta\rho = -\tilde{\mathbf{V}}_{\mathrm{ss}} \bigg/ \nabla \cdot \left(\frac{1}{\rho_0^2}\nabla\mathbf{u}_0\right). \tag{36}
\label{eq:bulk_density_correction}
\end{equation}
Comparing the two different parameterizations of the acoustic wave equation---the scattering virtual source expressions Eq.~\eqref{eq:scattering_virtual_source} and Eq.~\eqref{eq:scattering_virtual_source_bulk}, the radiation patterns of different model parameters, and their correction formulas Eqs.~\eqref{eq:velocity_correction}-\eqref{eq:density_correction} versus Eqs.~\eqref{eq:bulk_correction}-\eqref{eq:bulk_density_correction}---it can be seen that the parameterization in terms of bulk modulus \(\kappa\) and density \(\rho\) is more conducive to eliminating the cross-talk between bulk modulus \(\kappa\) and density \(\rho\). This is because the coherence between the radiation patterns generated by the \(\kappa\) and \(\rho\) parameterization is poorer than that between the radiation patterns generated by the \(\mathbf{v}\) and \(\rho\) parameterization.

If density variations are neglected, the acoustic wave equation Eq.~\eqref{eq:acoustic_wave} reduces to the scalar wave equation:
\begin{equation}
\left\{\frac{1}{\mathbf{v}^2(\mathbf{x})}\frac{\partial^2}{\partial t^2} - \nabla^2\right\} \mathbf{u}(\mathbf{x},t;\mathbf{x}_s) = \mathbf{s}(\mathbf{x}_s,t). \tag{37}
\label{eq:scalar_wave}
\end{equation}
From the acoustic PSDFWI above, the correction formula for velocity \(\mathbf{v}\) can be obtained:
\begin{equation}
\delta\mathbf{v} = \tilde{\mathbf{V}}_{\mathrm{ss}} \bigg/ \frac{2}{\mathbf{v}_0^3}\frac{\partial^2\mathbf{u}_0}{\partial t^2}. \tag{38}
\label{eq:scalar_velocity_correction}
\end{equation}
For comparison, the velocity correction formula in the steepest-descent FWI method is:
\begin{equation}
\delta\mathbf{v} = \tilde{\mathbf{V}}_{\mathrm{ss}} \frac{2}{\mathbf{v}_0^3}\frac{\partial^2\mathbf{u}_0}{\partial t^2}. \tag{39}
\label{eq:steepest_descent_correction}
\end{equation}
Comparing Eqs.~\eqref{eq:scalar_velocity_correction} and \eqref{eq:steepest_descent_correction}, the only difference is deconvolution versus cross-correlation. Furthermore, Eq.~\eqref{eq:steepest_descent_correction} also involves linearization under the Born approximation in the computation of the objective function gradient.

PSDFWI does not require the computation of the so-called objective function gradient, nor the Hessian matrix or its inverse for multi-parameter decoupling. Compared to current full-waveform inversion methods, it has the characteristics of lower computational cost and lower complexity, representing an innovative full-waveform inversion method.

The core steps of PSDFWI have been validated in previous work through numerical experiments on the Marmousi model \cite{chen2016b}, demonstrating its feasibility in velocity FWI.

\section{Physical-Structure-Driven Stratigraphic Physical Properties Imaging and Angle-Domain Common-Image Gather Inversion (PSDSI)}

Given a smooth subsurface medium model with accurate seismic kinematic characteristics, using the primary reflection wavefield in the observed wavefield, PSDSI comprises two components: (1) fast imaging of stratigraphic physical property variations (relative impedance perturbations), producing angle-independent (also called angle-domain averaged) images of stratigraphic physical property variations; (2) during the stratigraphic physical properties imaging process, angle decomposition of the wavefield is performed to obtain reflection opening angle-domain common-image gathers, followed by linear inversion of the angle-domain common-image gathers, utilizing the redundancy in the angle domain to achieve fine imaging of stratigraphic physical property (lithologic) parameters.

\subsection{Fast Imaging of Stratigraphic Physical Property Variations (Relative Impedance Perturbations)}

According to the PSDWII framework above, given a smooth subsurface medium model \(\mathbf{m}_s\) with accurate seismic kinematic characteristics, the steps for fast stratigraphic physical properties imaging using the primary reflection wavefield are:

\begin{enumerate}
\item Construct the subsurface incident wavefield;
\item Reverse-time extrapolate the observed primary reflection wavefield to reconstruct the subsurface reflection wavefield, i.e., obtain the approximate inversion of the subsurface volume reflection virtual source based on the relative perturbation of model parameters;
\item Based on the theoretical expression of the subsurface volume reflection virtual source, use the inverted subsurface volume reflection virtual source and the subsurface incident wavefield to image the subsurface stratigraphic physical property variations.
\end{enumerate}

Unlike methods that obtain impedance images by integrating reflectivity images, the physical-structure-driven wave-equation-based stratigraphic physical properties imaging directly utilizes the theoretical expression of the subsurface reflection virtual source derived from the wave equation, obtaining stratigraphic physical property images through inversion. Its data requirements, computational conditions, and computational efficiency are comparable to RTM, and it does not rely on the horizontally layered medium assumption.

Given smooth velocity and density models with accurate seismic kinematic characteristics (\(\mathbf{v}_s\) and \(\rho_s\)), for the acoustic wave equation Eq.~\eqref{eq:acoustic_wave}, the specific computational formulas for stratigraphic physical properties imaging are:

1) Construct the subsurface incident wavefield according to Eq.~\eqref{eq:incident_wave}:
\begin{equation}
\mathbf{u}_0 = \mathbf{G}_{\mathrm{i}}(\mathbf{v}_s,\rho_s)\mathbf{s}. \tag{40}
\label{eq:incident_wave_imaging}
\end{equation}

2) According to Eq.~\eqref{eq:approx_virtual_source_primary}, obtain the approximate inversion of the primary volume reflection virtual source based on the relative perturbation of model parameters (i.e., reconstruction of the subsurface reflection wavefield):
\begin{equation}
\tilde{\mathbf{V}}_{\mathrm{sr}}^{\mathrm{p}} = \mathbf{G}_{\mathrm{s}}^{*}(\mathbf{v}_s,\rho_s)\delta\mathbf{u}^{\mathrm{p}}. \tag{41}
\label{eq:approx_virtual_source_reflection}
\end{equation}
Unlike PSDFWI, which uses the wavefield residual at the receivers to invert the scattering virtual source, here \(\delta\mathbf{u}^{\mathrm{p}}\) is the observed primary reflection wavefield. In essence, the wavefield on the right-hand side of Eq.~\eqref{eq:approx_virtual_source_reflection} is also the wavefield residual between the observed wavefield and the computed wavefield in the smooth model. For the surface-source and surface-receiver acquisition system, the computed wavefield at the receivers is zero, so the wavefield residual equals \(\delta\mathbf{u}^{\mathrm{p}}\).

3) According to the theoretical expression of the virtual source Eq.~\eqref{eq:volume_reflection_virtual_source_primary}, using \(\tilde{\mathbf{V}}_{\mathrm{sr}}^{\mathrm{p}}\) and \(\mathbf{u}_0\), obtain the angle-domain averaged relative impedance perturbation \(\overline{\mathbf{I}}_{\mathrm{r}}(\alpha_{\rho},\alpha_{\mathrm{v}})\):
\begin{equation}
\overline{\mathbf{I}}_{\mathrm{r}}(\alpha_{\rho},\alpha_{\mathrm{v}}) = \tilde{\mathbf{V}}_{\mathrm{sr}}^{\mathrm{p}} \bigg/ \left(\frac{1}{\rho_s\mathbf{v}_s^2}\frac{\partial^2\mathbf{u}_0}{\partial t^2}\right). \tag{42}
\label{eq:avg_impedance_perturbation}
\end{equation}
where \(\overline{\mathbf{I}}_{\mathrm{r}}(\alpha_{\rho},\alpha_{\mathrm{v}}) = 2\alpha_{\rho} + \alpha_{\mathrm{v}} = 2\frac{\rho_s\delta\mathbf{v} + \mathbf{v}_s\delta\rho}{\rho_s\mathbf{v}_s} = 2\frac{\delta\mathbf{I}}{\mathbf{I}_s}\).

The above stratigraphic physical properties imaging method can quickly obtain property variation information, but it reflects the combined variation of stratigraphic physical properties, containing coupling effects between multiple parameters, which is not conducive to detailed interpretation of stratigraphic physical property variations. If, during the stratigraphic physical properties imaging process, the wavefields are decomposed in the angle domain to obtain angle-domain common-image gathers, we know that the coupling effects between multiple parameters vary with the angle between the incident and reflected waves---i.e., the radiation pattern of the reflection virtual source varies with the reflection opening angle. Therefore, the redundant information in the angle domain of angle-domain common-image gathers provides a path for decoupling the coupling effects between multiple parameters.

\subsection{Angle-Domain Common-Image Gather Inversion for Stratigraphic Physical Properties Imaging}

According to the PSDWII framework above, the steps for angle-domain common-image gather inversion in stratigraphic physical properties imaging are:

\begin{enumerate}
\item Construct the subsurface incident wavefield and perform angle-domain decomposition;
\item Reconstruct the subsurface reflection wavefield (inversion of the subsurface reflection virtual source) and perform angle-domain decomposition;
\item Based on the theoretical expression of the subsurface volume reflection virtual source, use the inverted subsurface volume reflection virtual source and the subsurface incident wavefield to image the subsurface stratigraphic physical property variations, producing angle-domain common-image gather data;
\item Using the theoretical expression of the volume reflection virtual source and the redundant information in the angle domain, perform linear inversion on the angle-domain common-image gather data to obtain fine images of stratigraphic physical property parameters. Then, based on petrophysics, convert the stratigraphic physical property images into stratigraphic lithology images.
\end{enumerate}

Unlike those angle-domain common-image gather inversion methods that use the Zoeppritz equations or their approximations, PSDSI's angle-domain common-image gather inversion directly utilizes the theoretical expression of the subsurface reflection virtual source derived from the wave equation, which is based on the relative perturbation of stratigraphic physical property parameters varying with the reflection opening angle. It does not rely on the plane-wave or horizontally layered medium assumptions.

Given smooth velocity and density models with accurate seismic kinematic characteristics (\(\mathbf{v}_s\) and \(\rho_s\)), for the acoustic wave equation Eq.~\eqref{eq:acoustic_wave}, the specific computational formulas for angle-domain common-image gather inversion in stratigraphic physical properties imaging are:

1) Construct the subsurface incident wavefield according to Eq.~\eqref{eq:incident_wave}:
\begin{equation}
\mathbf{u}_0 = \mathbf{G}_{\mathrm{i}}(\mathbf{v}_s,\rho_s)\mathbf{s}. \tag{43}
\label{eq:incident_wave_angle}
\end{equation}
Perform angle decomposition on the incident wavefield \(\mathbf{u}_0\) to obtain \(\mathbf{u}_0(\theta)\).

2) According to Eq.~\eqref{eq:approx_virtual_source_primary}, obtain the approximate inversion of the primary volume reflection virtual source based on the relative perturbation of model parameters (i.e., reconstruction of the subsurface reflection wavefield):
\begin{equation}
\tilde{\mathbf{V}}_{\mathrm{sr}}^{\mathrm{p}} = \mathbf{G}_{\mathrm{s}}^{*}(\mathbf{v}_s,\rho_s)\delta\mathbf{u}^{\mathrm{p}}. \tag{44}
\label{eq:approx_virtual_source_angle}
\end{equation}
Perform angle decomposition on the inverted virtual source \(\tilde{\mathbf{V}}_{\mathrm{sr}}^{\mathrm{p}}\) to obtain \(\tilde{\mathbf{V}}_{\mathrm{sr}}^{\mathrm{p}}(\theta)\).

3) According to the theoretical expression of the subsurface volume reflection virtual source Eq.~\eqref{eq:volume_reflection_virtual_source_primary}, using the angle-domain \(\tilde{\mathbf{V}}_{\mathrm{sr}}^{\mathrm{p}}\) and the angle-domain \(\mathbf{u}_0\), image the subsurface stratigraphic physical property variations to obtain the angle-domain common-image gather data \(\tilde{\mathbf{I}}_{\mathrm{r}}(\theta)\) of \(\mathbf{I}_{\mathrm{r}}(\alpha_{\rho},\alpha_{\mathrm{v}},\theta)\):
\begin{equation}
\tilde{\mathbf{I}}_{\mathrm{r}}(\theta) = \tilde{\mathbf{V}}_{\mathrm{sr}}^{\mathrm{p}}(\theta) \bigg/ \left(\frac{1}{\rho_s\mathbf{v}_s^2}\frac{\partial^2\mathbf{u}_0(\theta)}{\partial t^2}\right). \tag{45}
\label{eq:angle_cig_data}
\end{equation}

4) Using the theoretical expression of \(\mathbf{I}_{\mathrm{r}}(\alpha_{\rho},\alpha_{\mathrm{v}},\theta)\) and the redundant information in \(\tilde{\mathbf{I}}_{\mathrm{r}}(\theta)\) in the angle domain, invert the stratigraphic physical property parameters:
\begin{equation}
\mathbf{I}_{\mathrm{r}}(\alpha_{\rho},\alpha_{\mathrm{v}},\theta) = (1 + \cos\theta)\alpha_{\rho} + \alpha_{\mathrm{v}} = \tilde{\mathbf{I}}_{\mathrm{r}}(\theta). \tag{46}
\label{eq:angle_inversion}
\end{equation}
Solve the system of linear equations formed by different \(\theta\) values to obtain the relative perturbations of velocity and density, \(\alpha_{\mathrm{v}}\) and \(\alpha_{\rho}\). Based on petrophysics, the relative perturbations of velocity and density can be further converted into relative perturbations of lithologic parameters.

The fast imaging and angle-domain inversion workflows of PSDSI have been validated in previous work through numerical experiments on acoustic random layered models and Marmousi model \cite{liu2020}, confirming their effectiveness in inverting stratigraphic physical property parameters.

\section{Physical-Structure-Driven Stratigraphic Structures Imaging Method (PSDMig)}

PSDMig is the imaging of stratigraphic boundaries, mathematically achieved by imaging the local reflectivity of stratigraphic boundaries using the primary reflection wavefield in the observed wavefield. Given a smooth subsurface medium model \(\mathbf{m}_s\) with accurate seismic kinematic characteristics, according to the PSDWII framework above, the implementation steps of the stratigraphic structures imaging method are:

\begin{enumerate}
\item Construct the subsurface incident wavefield;
\item Reverse-time extrapolate the observed primary reflection wavefield to reconstruct the subsurface reflection wavefield, i.e., obtain the approximate inversion \(\tilde{\mathbf{V}}_{\mathrm{sr}}^{\mathrm{p}}\) of \(\mathbf{V}_{\mathrm{sr}}^{\mathrm{p}}(\mathbf{m}_s,r)\);
\item Based on the theoretical expression of \(\mathbf{V}_{\mathrm{sr}}^{\mathrm{p}}(\mathbf{m}_s,r)\), use the constructed subsurface incident wavefield \(\mathbf{u}_0\) and \(\tilde{\mathbf{V}}_{\mathrm{sr}}^{\mathrm{p}}\) to image the local reflectivity \(r\) of the stratigraphic boundary, thereby achieving stratigraphic structures imaging.
\end{enumerate}

Unlike Claerbout's migration imaging principle \cite{claerbout1971}, where the wavefield extrapolation formula originates from the wave equation for heterogeneous media and the wavefield imaging formula originates from the simplified Zoeppritz equations for horizontally layered media, PSDMig directly utilizes the theoretical expression of the subsurface reflection virtual source based on the local reflectivity of the stratigraphic boundary and the primary reflection wave equation, both derived from the wave equation, to achieve imaging of the local reflectivity of stratigraphic boundaries through inversion, i.e., images of stratigraphic structure.

Given smooth velocity and density models with accurate seismic kinematic characteristics (\(\mathbf{v}_s\) and \(\rho_s\)), for the acoustic wave equation Eq.~\eqref{eq:acoustic_wave}, the specific computational formulas for PSDMig are:

1) Construct the subsurface incident wavefield according to Eq.~\eqref{eq:incident_wave}:
\begin{equation}
\mathbf{u}_0 = \mathbf{G}_{\mathrm{i}}(\mathbf{v}_s,\rho_s)\mathbf{s}. \tag{47}
\label{eq:incident_wave_mig}
\end{equation}

2) Using Eq.~\eqref{eq:approx_virtual_source_primary}, obtain the approximate inversion \(\tilde{\mathbf{V}}_{\mathrm{sr}}^{\mathrm{p}}\) of \(\mathbf{V}_{\mathrm{sr}}^{\mathrm{p}}(\mathbf{v}_s,\rho_s,r)\) (i.e., reconstruction of the subsurface reflection wavefield):
\begin{equation}
\tilde{\mathbf{V}}_{\mathrm{sr}}^{\mathrm{p}} = \mathbf{G}_{\mathrm{s}}^{*}(\mathbf{v}_s,\rho_s)\delta\mathbf{u}^{\mathrm{p}}. \tag{48}
\label{eq:approx_virtual_source_mig}
\end{equation}

3) According to the theoretical expression Eq.~\eqref{eq:surface_reflection_virtual_source_primary} of \(\mathbf{V}_{\mathrm{sr}}^{\mathrm{p}}(\mathbf{v}_s,\rho_s,r)\), using \(\tilde{\mathbf{V}}_{\mathrm{sr}}^{\mathrm{p}}\) and the constructed \(\mathbf{u}_0\), image the local reflectivity \(r\) of the stratigraphic boundary, thereby achieving stratigraphic structures imaging:
\begin{equation}
r = -\tilde{\mathbf{V}}_{\mathrm{sr}}^{\mathrm{p}} \bigg/ \left(\frac{1}{\rho_s}\frac{\partial\mathbf{u}_0}{\partial t}\right). \tag{49}
\label{eq:local_reflectivity_imaging}
\end{equation}

For the scalar wave equation Eq.~\eqref{eq:scalar_wave}, Eq.~\eqref{eq:local_reflectivity_imaging} reduces to:
\begin{equation}
r = -\tilde{\mathbf{V}}_{\mathrm{sr}}^{\mathrm{p}} \bigg/ \left(\frac{1}{\mathbf{v}_s}\frac{\partial\mathbf{u}_0}{\partial t}\right). \tag{50}
\label{eq:local_reflectivity_scalar}
\end{equation}

The comparison between the core imaging formula of PSDMig and Claerbout's cross-correlation imaging condition, as well as imaging tests on point scatterers and reflecting surfaces, have been completed in previous work \cite{chen2016a}.

\section{Discussion}

\subsection{Essential Difference Between Physical-Structure-Driven and Mathematical Optimization Driven Approaches}

There is a fundamental methodological difference between the PSDWII framework and the current mainstream mathematical optimization driven approaches. The mainstream approaches formulate seismic waveform inversion imaging as a least-squares optimization problem: constructing an objective function between observed and modeled data, computing the gradient of the objective function with respect to the model parameters, and updating the model through iterative descent algorithms. The core of this paradigm is mathematical optimization theory, where the physical process of seismic wave propagation is embedded in forward modeling, but the inversion process itself is dominated by mathematical logic.

The PSDWII framework, on the other hand, starts from the physical structure of seismic wave propagation, decomposing the propagation process into three physical stages: ``incident wave propagation $\rightarrow$ virtual source generation $\rightarrow$ secondary wave propagation.'' The inversion strategy is customized for each stage through physical driving: first, the adjoint operator (reverse-time extrapolation) is used to extract the approximate inversion of the virtual source from the observed wavefield; then, based on the theoretical expression of the virtual source, the incident wavefield and radiation pattern are removed; finally, the target parameters are extracted. This process is not driven by an objective function or guided by gradient descent, but follows the physical causal chain in reverse.

The core operational difference between the two paradigms can be more intuitively expressed as:

\textbf{Mathematical optimization driven by the gradient method:}
$$
\text{Model update or inversion target} \propto 
$$
\text{Cross-correlation of adjoint wavefield with wave operator derivative and incident wavefield}
$$
$$
\textbf{Physical-structure-driven (PSDWII):}
$$
\text{Model update or inversion target} \propto
$$
\text{Deconvolution of adjoint wavefield with wave operator derivative and incident wavefield}
$$
$$
The difference between the two paradigms can be summarized as: the mathematical optimization driven method answers ``how to make the modeled data consistent with the observed data,'' while the physical-structure-driven method answers ``what physical process in the subsurface does the observed wavefield originate from.'' The former is a mathematical approximation in data space, while the latter is a causal tracing in physical space.

\subsection{On the Problem of Multi-Parameter Coupling}

The coupling problem in multi-parameter inversion is one of the core difficulties faced by current FWI. The analysis in this paper shows that multi-parameter coupling consists of two levels: physical coupling and mathematical coupling.

Physical coupling is the phenomenon where different model parameters are naturally coupled together during wave propagation, jointly determining the characteristics of wavefield propagation. This is an objective physical reality and is also the source of rich information about the subsurface medium. Mathematical coupling is artificially introduced by the computation of the objective function gradient in the optimization framework. In gradient-based algorithms, not only is the coupling between multiple parameters not decoupled, but the coupling effect is further aggravated.

From the perspective of mathematical operations, the root cause of mathematical coupling in the gradient method lies in its cross-correlation operation. The virtual source, as the combined effect of the incident wavefield, wave operator derivative, and parameter perturbation, already has a coupled physical meaning. The gradient method further combines the adjoint wavefield with the incident wavefield and wave operator derivative through cross-correlation, adding mathematical coupling on top of the physical coupling, making the coupling between different parameters even more difficult to decouple. This understanding explains why gradient-based methods rely on the computationally expensive inverse of the Hessian matrix to decouple both the physical and mathematical couplings in the objective function gradient.

The PSDWII framework does not define an objective function, does not compute gradients, preserves physical coupling, and does not generate mathematical coupling. For the deconvolution of the virtual source with the wave operator derivative and incident wavefield, if the coherence between the radiation patterns of different parameters is poor, then the cross-talk term in Eq.~\eqref{eq:ct_definition} tends to zero, achieving decoupling of the physical coupling.

The above discussion indicates that model parameterization has an important influence on multi-parameter decoupling. Comparing two parameterizations of the acoustic wave equation---velocity and density versus bulk modulus and density---the bulk modulus and density parameterization is more conducive to eliminating cross-talk, because the coherence between the radiation patterns of bulk modulus and density is poorer than that between the radiation patterns of velocity and density.

\subsection{Theoretical Consistency Between PSDMig and Least-Squares Migration}

Based on the analysis of incident wavefield reconstruction and fidelity in Section 3, when the incident wavefield in the denominator of PSDMig is the wavefield reconstructed through combined forward and adjoint propagation, the propagation operators and adjoint propagation operators in the numerator and denominator of its core formula Eq.~\eqref{eq:local_reflectivity_imaging} approximately cancel under the condition of time consistency:
\begin{equation}
\frac{\mathbf{G}_{\mathrm{s}}^{*}(\mathbf{m}_0)\mathbf{G}_{\mathrm{s}}(\mathbf{m}_0)\mathbf{u}_0}{\mathbf{G}_{\mathrm{i}}^{*}(\mathbf{m}_0)\mathbf{G}_{\mathrm{i}}(\mathbf{m}_0)\mathbf{u}_0} \approx 1 \tag{51}
\label{eq:lsm_equivalence}
\end{equation}
This cancellation effect is mathematically equivalent to the role of the inverse of the Hessian matrix \( (\mathbf{G}^T\mathbf{G})^{-1} \) in least-squares migration (LSM). Therefore, PSDMig theoretically has the same goal as least-squares migration---obtaining amplitude-preserved, high-resolution reflectivity images.

However, PSDMig does not achieve this by solving a least-squares problem or computing the inverse of the Hessian matrix, but rather through physical-structure-driven deconvolution operations, accomplished naturally in an analytical manner. This makes PSDMig computationally more efficient than least-squares migration, while being more intuitive in terms of physical mechanism. The difference between the two is: least-squares migration is a mathematically driven compensation, while PSDMig is a physically driven compensation.

\subsection{Essential Difference Between PSDMig and FWI-Imaging}

The commonly used FWI-Imaging technique in the current industry follows this basic idea: first obtain a high-resolution, high-signal-to-noise ratio model through FWI, then perform directional derivative operations on the model to extract boundaries with sharp model variations as structure imaging results.

There are three levels of essential difference between PSDMig and FWI-Imaging:

(1) Difference in theoretical foundation: FWI-Imaging is essentially an edge detection technique in image processing, operating on the FWI inversion result rather than the seismic wavefield; PSDMig starts from the primary reflection approximation of the reflection wave equation and directly uses the primary reflection wavefield to invert the local reflectivity of reflector boundaries---a structure imaging method fully based on the wave equation.

(2) Dependence on FWI results: FWI-Imaging highly depends on the quality of the FWI result; if the FWI result has insufficient resolution or contains errors, the FWI-Imaging result is unreliable. PSDMig only requires a kinematically accurate smooth background model and does not require high-resolution FWI results, making it more feasible under real data conditions.

(3) Physical meaning of reflectivity definition: The directional derivative in FWI-Imaging is a purely mathematical operation lacking clear physical meaning. The local reflectivity \(r\) in PSDMig is defined along the incident wave propagation direction, with a clear physical causality---only when the incident wave scans property variations along its propagation direction can reflected waves be generated.

\subsection{Data and Model Requirements of the Three Methods}

The data, initial model, and computational requirements of the three specific methods in the PSDWII framework are comparable to those of current mainstream methods:

PSDFWI has data requirements consistent with conventional FWI, requiring broadband, wide-azimuth observed data; the initial model requirements are consistent with conventional FWI; computational conditions are wavefield forward modeling and reverse-time extrapolation, comparable to conventional FWI.

PSDSI has data requirements consistent with pre-stack AVO/AVA inversion, requiring amplitude-preserved primary reflection data; the imaging model requirement is a smooth subsurface medium model with accurate seismic kinematic characteristics, consistent with the migration velocity model requirement for RTM; computational conditions are wavefield forward modeling and reverse-time extrapolation plus angle decomposition, roughly comparable to RTM.

PSDMig has data requirements consistent with RTM, requiring primary reflection data; the imaging model requirement is a smooth subsurface medium model with accurate seismic kinematic characteristics, consistent with the migration velocity model requirement for RTM; computational conditions are wavefield forward modeling and reverse-time extrapolation, the same as RTM.

Under the condition of having a smooth subsurface medium model with accurate seismic kinematic characteristics, for isotropic elastic waves in PSDSI and PSDMig, P/S wave decomposition of the elastic wavefield is not required during the imaging process \cite{chen2024}.

This shows that the three methods in the PSDWII framework maintain the advantage of physical structure driving without adding extra data or computational burden.

\subsection{Previous Numerical Validation}

The core steps of the three methods in the PSDWII framework have been validated in our previous numerical experiments. PSDFWI experiments on the Marmousi model demonstrated that the time-second-order integration strategy effectively mitigates the cycle-skipping problem, and the deconvolution-type velocity correction formula achieved better inversion results than the cross-correlation-type formula \cite{chen2016b}. PSDSI experiments on acoustic random layered models andMarmousi model validated the effectiveness of fast imaging and angle-domain common-image gather inversion \cite{liu2020}. PSDMig imaging tests on point scatterers and reflecting surfaces showed that the imaging results are superior to those of Claerbout's cross-correlation imaging condition in terms of resolution and phase fidelity \cite{chen2016a}. The above numerical validations provide conceptual support for the theoretical framework of this paper and also demonstrate that the three methods of the PSDWII framework are feasible in actual computation.

\subsection{Tension Between Theory and Engineering Implementation}

The theoretical derivation of the PSDWII framework is based on the idealized assumptions of time consistency and first-order Born approximation (primary reflection approximation), which provide it with clear physical logic and mathematical expressions. However, just as the theoretical perfection of Newton's method does not directly translate to engineering feasibility, PSDWII also faces certain challenges in practical implementation.

The deconvolution (division) operation is analytical in theory, but under conditions of band-limited data, noise, and limited illumination in real data, its stability requires special numerical treatment. The analysis in Section 3 regarding the approximate cancellation of propagation operators and adjoint propagation operators has its accuracy dependent on the traveltime accuracy of the background velocity model; the reconstruction quality of the incident wavefield in the denominator is constrained by the acquisition system and data frequency band. These factors together determine the effective boundary of the PSDWII method in practical applications.

This recognition does not diminish the theoretical value of the PSDWII framework, but rather identifies several key technical issues that need to be addressed for its transition from theory to engineering application. This is one of the key directions for future research on the PSDWII framework, and also the necessary path from theoretical innovation to practical application. Just as the engineering practicality of the gradient method does not negate the theoretical superiority of Newton's method, the challenges in engineering implementation of the PSDWII framework do not affect its value as a physical-structure-driven inversion imaging methodology.

\section{Conclusions}

\begin{enumerate}
\item The PSDWII framework proposed in this paper transforms the conventional ``mathematical optimization driven'' paradigm of waveform inversion imaging into a ``physical-structure-driven'' paradigm, shifting seismic waveform inversion imaging research from using general mathematical optimization methods---with poor physical interpretability and high computational complexity---to tailoring physical-structure-driven inversion imaging methods according to the physical structure of the seismic waveform inversion imaging problem---with strong physical interpretability and computational conciseness and efficiency.

\item Seismic wave propagation has a clear physical structure---``two propagations, one virtual source'': incident wave propagation $\rightarrow$ interaction of incident wave with heterogeneities generating virtual sources with radiation patterns, which excite secondary waves $\rightarrow$ secondary wave propagation and reception. This physical structure is the physical foundation of the PSDWII framework.

\item The core logic of the PSDWII framework is the three-step structure of ``adjoint projection $\rightarrow$ deconvolution removal $\rightarrow$ target parameter extraction.'' The adjoint wavefield is explicitly interpreted as the approximate inversion of the subsurface virtual source, endowing the adjoint wavefield with a clear physical meaning. The time consistency between the virtual source and the incident wavefield is a necessary condition for determining whether the adjoint wavefield is the true virtual source.

\item The core operation of the PSDWII framework is deconvolution removal (division), which deconstructs and separates the incident wavefield and radiation pattern from the virtual source, directly extracting pure physical property parameters. In contrast, the core operation of the traditional gradient method is cross-correlation, which permanently mixes the contributions of different parameters in the gradient. This fundamental difference enables PSDWII to completely avoid the generation of mathematical coupling at the methodological level, eliminating the need for compensatory mathematical decoupling means such as Hessian inversion.

\item Based on the PSDWII framework, three physical-structure-driven specific methods are constructed: the physical-structure-driven full-waveform inversion method (PSDFWI), the physical-structure-driven stratigraphic physical properties imaging and angle-domain common-image gather inversion method (PSDSI), and the physical-structure-driven stratigraphic structures imaging method (PSDMig). The three methods share a unified methodological core, adopting different virtual source expressions, and respectively serve the three levels of inversion targets: model building, lithologic parameter inversion, and structure boundary imaging. The data, initial model, and computational requirements of the three methods are comparable to those of current mainstream methods, without adding extra burden.

\item The PSDWII framework only involves physical coupling and does not introduce mathematical coupling. It does not construct objective functions, nor compute objective function gradients or Hessian matrices and their inverses. By utilizing the angular dependence differences of the radiation patterns of different model parameters, physically driven decoupling of multiple parameters can be achieved. The choice of model parameterization has an important influence on the effectiveness of multi-parameter decoupling.

\item The local reflectivity is defined along the incident wave propagation direction, applicable to arbitrary incidence angles and arbitrary complex structures, with clear physical causality. The traditional definition along the interface normal direction is only a special case. The PSDMig method based on this definition is superior to the FWI-Imaging method based on image processing in terms of physical essence and practical feasibility. PSDMig theoretically has the same goal as least-squares migration, but is implemented through physical-structure-driven deconvolution operations, with computational efficiency superior to least-squares migration.
\end{enumerate}

\end{document}